\documentclass[allclo]{FBSart}
\usepackage{amsfonts}
\usepackage{amssymb}

\title{Structure of $A$=3 Nuclear Systems Using Realistic Hamiltonians}
\author{L. E. Marcucci\instnr{1,2}\thanks{\textit{E-mail address:} 
laura.marcucci@df.unipi.it}, 
L. Girlanda\instnr{2}, A. Kievsky\instnr{2}, S. Rosati\instnr{1,2}, 
M.\ Viviani\instnr{2}}
\instlist{Department of Physics ``E. Fermi'',  
University of Pisa, I-56127 Pisa, Italy
\and 
INFN-Sezione di Pisa, I-56127 Pisa, Italy}

\runningauthor{L.\,E.\,Marcucci et al.}
\runningtitle{Structure of Light Nuclear Systems Using Realistic Hamiltonians}
\begin{document}

\maketitle
\begin{abstract}
The structure of $A$=3 low-energy scattering states 
is described using the hyperspherical harmonics 
method with realistic Hamiltonian models, consisting of two- and three-nucleon 
interactions. Both coordinate and momentum space two-nucleon potential 
models are considered. 
\end{abstract}

\section{Introduction}
\label{sec:intro}

One of the main ingredients necessary to study few-body nuclear systems
is a realistic description of the nuclear interaction. A number of 
nucleon-nucleon ($NN$) potentials has been determined in the recent 
years. They all reproduce the deuteron binding energy and 
fit a large set of $NN$ scattering data below the pion-production 
threshold with a $\chi^2$/datum of about 1. Among these potentials, 
we will consider in the present study only the ``phenomenological'' model 
of Ref.~\cite{Wir95} (AV18),  and a model
based on chiral symmetry derived in Ref.~\cite{Ent03} (N3LO-Idaho). 
Among the many features of these two models, we note only that
the AV18 is a local $NN$ potential model, with a strong short-range 
repulsion and tensor component, while the N3LO-Idaho is a non-local 
$NN$ potential model, with a softer short-range repulsion and tensor component 
than the AV18. As a consequence of these differences, 
it is interesting to test these potential models 
studying light nuclear systems. In these systems, a further 
contribution to the realistic nuclear Hamiltonian model
comes from the three-nucleon 
interaction (TNI). Several models of TNI's have been proposed. They are  
mainly based on the exchange of pions among the three 
nucleons, as the Urbana IX (UIX) TNI~\cite{Pud97}, which 
will be considered in the present study. The more recent TNI 
models studied within the chiral approach~\cite{Epe02} 
and the extension of the UIX model known as the Illinois 
TNI~\cite{Pie01} will 
be considered in a near future.

A second crucial ingredient in the study of light nuclear systems is the 
technique used to solve the $A$-body Schr\"odinger equation. Several 
methods have been developed in the past years (see 
Refs.~\cite{Glo96,Car98} for a review). Among them, we consider in the 
present study the technique known as the hyperspherical harmonics (HH) 
method, which will be briefly described in the following section.
In Sec.~\ref{sec:res}, the 
results for the $n-d$ and $p-d$ scattering lengths will be 
presented and 
compared with the available experimental data.

\section{The Hyperspherical Harmonics Method}
\label{sec:hhmethod}

The nuclear wave function for an $A$-body system can be written as 
\begin{equation}
|\Psi\rangle=\Sigma_\mu c_\mu |\Phi_\mu\rangle \ ,
\label{eq:psi}
\end{equation}
where $|\Phi_\mu\rangle$ is a suitable complete set of 
states, and $\mu$ is an index denoting the set of quantum numbers 
necessary to completely determine the basis elements. 
In the present work, the functions $|\Phi_\mu\rangle$ have been written 
in terms of HH functions 
both in configuration-space or in momentum-space~\cite{Viv06}.
The unknown coefficients $c_\mu$ of Eq.~(\ref{eq:psi}) are obtained 
applying the 
Rayleigh-Ritz (Kohn) variational principle for the bound (scattering) state 
problem. Then, the matrix elements of the different operators of  
the Hamiltonian are calculated, working in coordinate- or in 
momentum-space depending on what is more convenient. Thus, 
the problem is reduced to an eigenvalue-eigenvector 
problem (system of algebraic linear equations), 
which can be solved with standard numerical techniques~\cite{Kie97-98}.

\section{Results}
\label{sec:res}

The $n-d$ and $p-d$ doublet and quartet scattering lengths 
obtained with  
the non-local N3LO-Idaho~\cite{Ent03} $NN$ interaction, with or 
without the inclusion of the UIX TNI~\cite{Pud97}, are given in 
Table 1,
and compared with the available experimental data ~\cite{Dil71}.
Also shown are the results obtained with the local AV18~\cite{Wir95} 
$NN$ interaction and the AV18/UIX potential model for a 
comparison~\cite{Kie95}.
Note that in the case of the N3LO-Idaho/UIX model, the parameter 
in front of the spin-isospin independent part of the UIX TNI has 
been rescaled by a factor of 0.384 to fit the triton binding energy. 
In this way, the triton, $^3$He, and $^4$He binding energies are
8.481 MeV, 7.730 MeV, and 28.534 MeV, respectively. 
Furthermore, the N3LO-Idaho and 
N3LO-Idaho/UIX results shown in the table are accurate at the 
10$^{-3}$ fm level. In fact, the convergence of the HH expansion 
has been tested with a procedure similar to the one used in 
Ref.~\cite{Viv06} for the $A$=3 and 4 bound states observables.

From inspection of the table we can conclude that: 
(i) both the $n-d$ and $p-d$ quartet scattering lengths are very 
little model-dependent. Also, they are not affected by the inclusion 
of the TNI. The trend shown by the  
AV18 and AV18/UIX results has been found also  
in the case of the non-local N3LO-Idaho and 
N3LO-Idaho/UIX potential models. 
(ii) The $n-d$ doublet scattering length is very 
sensitive to the choice of the $NN$ potential model, when no TNI is included. 
However, once the TNI is included, and therefore the triton 
binding energy is well reproduced, $^2a_{nd}$ becomes model-independent. 
This is a well-known 
feature, related to the fact that $^2a_{nd}$ and the triton binding energy
are linearly correlated (the so-called 
Phillips line~\cite{Phi77}). (iii) The $p-d$ doublet scattering length 
is positive and quite model-dependent, if only the two-nucleon interaction
is included. Once the TNI is added, $^2a_{pd}$ becomes very little and 
negative. Some model-dependence remains, but the problem of extrapolating 
to zero energy the experimental results makes impossible any 
meaningful comparison between theory and experiment.

In conclusion, the application of the HH method to treat the 
low-energy scattering problem using non-local $NN$ interactions 
has been found successful. Both $n-d$ and $p-d$ systems have been 
considered, with the full inclusion of the Coulomb interaction, in the 
second case. A similar investigation for the $A$=4 scattering 
lengths has been reported in Ref.~\cite{Viv07}. Further work at higher 
energies is currently underway.

\begin{table}[hbt]
\beforetab
\begin{tabular}{cccccc}
\firsthline
 & AV18 & AV18/UIX 
& N3LO-Idaho & N3LO-Idaho/UIX & Exp. \\
\midhline
$^2a_{nd}$ (fm) 
& 1.27 & 0.63 & 1.100 & 0.623 & 0.65(4) \\
$^4a_{nd}$ (fm) 
& 6.33 & 6.33 & 6.342 & 6.343 & 6.35(2) \\
\midhline
$^2a_{pd}$ (fm) 
& 1.17 & -0.02 & 0.862 & -0.007 &  \\
$^4a_{pd}$ (fm) 
& 13.6 & 13.7  & 13.646 & 13.647 &  \\
\lasthline
\label{tab:nd}
\end{tabular}
\aftertab
\captionaftertab[]{Doublet and quartet $n-d$ and $p-d$ scattering lengths, 
obtained 
with different potential models. The experimental $n-d$ 
scattering lengths are taken from Ref.~\protect{\cite{Dil71}}, while the 
AV18 and AV18/UIX results are taken from 
Ref.~\protect\cite{Kie95}.}
\end{table}

\end{document}